\documentclass[a4paper,10pt]{article}
\usepackage[utf8]{inputenc}
\def\dis{\displaystyle}
\def\del{\partial}

\def\d{\rm d}

\title{Constraints on Operator Ordering from Third Quantization }
\author{ Yoshiaki Ohkuwa$^1$, Mir Faizal$^2$, Yasuo Ezawa$^3$\\ \\
$^1$Division of Mathematical Science, \\ Department of Social Medicine,\\ Faculty
of Medicine,
University of Miyazaki, \\ Kihara 5200,  Kiyotake-cho, Miyazaki, 
889-1692, Japan\\  \\
$^2$Department of Physics and Astronomy, \\  University of Waterloo,   Waterloo,\\
Ontario N2L 3G1, Canada\\ \\  
$^3$Department of Physics, Ehime university,\\ 2-5 Bunkyo-cho, Matsuyama, 
790-8577,
Japan }
\date{}
\begin{document}

\maketitle

\begin{abstract}
In this paper, we   analyse the Wheeler-DeWitt equation in the third quantized formalism. 
 We will demonstrate that for certain operator ordering,  the 
early stages of  the universe are dominated by quantum fluctuations, and the universe becomes classical at 
later stages during the cosmic expansion. This is physically expected,
if the universe is formed from quantum fluctuations in the third 
quantized formalism. So, we will argue that this physical requirement can be used 
to constrain the form of the operator ordering chosen. We will explicitly demonstrate this to be the case for two 
different   cosmological models.
\end{abstract}

\vspace{5mm}
\quad PACS numbers : 04.60.Ds, 04.60.Kz, 98.80.Qc
\vspace{3mm}

\section{Introduction}
The wave function of the universe contains all the physical information about the universe because 
it describes the quantum state of the universe 
  \cite{Hartle83}-\cite{Hawking}. In the no-boundary proposal the 
  wave function of the universe is obtained by summing over all  
four geometries and    field 
  configurations that match a specific field configuration on a spatial section. 
The wave function can also be obtained as a solution to the Wheeler-DeWitt equation, which 
can be viewed as the $\rm Schr\ddot{o}dinger$'s equation for gravity \cite{DeWitt67}-\cite{Wheeler57}.
However, just as the single particle $\rm Schr\ddot{o}dinger$'s equation 
cannot be used to analyse a  multi-particle 
system in the first quantized formalism, the  Wheeler-DeWitt equation cannot be used to describe multi-universe 
system in the second quantized formalism. However, a multi-particle system can be analysed by using 
 second quantization.
 This observation has motivated the study of third quantization of the Wheeler-DeWitt equation \cite{th}-\cite{th1}. 
 In the third quantized formalism, the Wheeler-DeWitt equation is viewed as a classical  field equation. 
 Thus, an action is constructed such that the 
 field equations corresponding to  that action are the Wheeler-DeWitt equation. This action describes the theory that
 can be third quantized. The third quantization of this action produces a multi-geometry theory. If these geometries 
 are identified with individual universes, then the third quantization of this action describes a multiverse
 \cite{mult}-\cite{mult1}. 
 
 The third quantization has also been used for analysing the virtual black holes   \cite{bh}. 
Here the fluctuations of the spacetime at Planck scale cause the formation of virtual black holes. 
This model of virtual black holes can be used to solve the problem of time. This is because in this model 
the entropy of the universe keeps increasing due to the interaction of these virtual black holes with matter. 
The direction of time can then be identified with the increase of the total entropy of the universe. 
The model for the spacetime foam can also be used to explain the end stage of the evaporation of real black holes
\cite{vbh}. 
In this model, real black holes evaporate down to Planck size and 
then disappear in the sea of virtual black holes. It may be noted that the third quantization has also been used 
to address the cosmological constant problem using the idea of baby universes \cite{ba}. 
In this model, the creation or annihilation of baby of universe in the third quantized formalism is 
similar to the creation or annihilation of a particle in the second quantized formalism. The propagator 
of the theory corresponded to a wormhole, and the third quantized version of the momentum conservation 
is represented by the conservation of the axion charge. 
 
 It is possible for the universe to form from quantum fluctuations in the third quantized formalism \cite{ab}. 
 Thus, it is expected that the early stages in the evolution of the universe would be 
dominated by  quantum fluctuations. It is known that the geometry of spacetime is described by  
a classical spacetime at later  stages, so it is expected that the 
quantum fluctuations will get minimized  at later stages of the cosmic expansion. 
 The uncertainty for a model of third quantized universes has been studied, and it  was observed that 
 the fluctuations in the third quantized formalism decrease very rapidly during the course of cosmic expansion  
 \cite{un}-\cite{un1}. 
 This uncertainty has also been studied for third quantized 
   Brans-Dicke theories  \cite{ai}. In  this analysis, the
  distribution function for the universes    has 
  been obtained.   The uncertainty principle has also been discussed in the context of the third quantization
   of $f(R)$ gravity theories \cite{f}-\cite{f1}. The distribution function for the universes in the 
  third quantization   Kaluza-Klein theories     
have also been obtained  \cite{ia}. 
  In this analysis, it  was   demonstrated that the compactification of 
  geometries is consistent with third quantization. The third quantization has been used to 
  analyse the quantum transitions from the string perturbative vacuum to cosmological 
  configurations which is characterized by isotropic contraction and decreasing dilaton \cite{st}.
  It was observed that such    transitions could be  represented by the 
  production of pairs of universes from the  vacuum state. All this analysis was done for specific choice 
  of operator ordering. However, it is known that operator ordering can have direct physical consequences  
  \cite{oo}-\cite{o1}. So, this motivates us to study the effect of operator ordering on the third 
  quantization, and this is what will be done in this paper.  We would like to point out that the main aim of this paper is to 
  use the physical requirements on cosmological models to restrict the form of operator ordering used. 
  This is because there is no mathematical way to prefer one choice of factor ordering from another. 
  
In this paper, we study the effect of operator ordering in the 
third quantized formalism. We will use the third quantized formalism for analysing the effect of uncertainty relation 
on the structure of spacetime during cosmic expansion. We will also discuss the operator ordering for this theory. 
We will observe that for a specific choice of operator ordering  
quantum fluctuations dominate at the early stage of the universe and the spacetime becomes classical at 
the later stages. 
This is physically expected, if the universe is formed from quantum fluctuations in the third 
quantized formalism. 
The remaining of the paper is organized as follows. In section \ref{1},  we will analyse the third quantization 
of  general relativity with a cosmological constant. In section \ref{2}, we will then study the uncertainty 
relation for this 
model, and in section  \ref{b}, we will analyse the operator ordering for this model. Then in section 
\ref{a}, we will apply this formalism for another minisuperspace model.
Finally, we will summarize our results in section \ref{d}. We will also suggest some possible 
extension of this works in this last section.

\section{Einstein Gravity}\label{1}
The universe is expected to form from the quantum fluctuations 
which could be described by 
 the third quantized theory \cite{ab}. 
So, we would need
to analyse the Wheeler-DeWitt equation in the third quantized formalism. So, in this section,  we will 
  analyse the third quantization of general relativity with a cosmological constant, 
since  we are interested in operator ordering problem 
which will be discussed precisely in section 4.
The action for general relativity with a 
  cosmological constant term is written as 
$$
S=\int \! d^4 x \sqrt{-g} {1 \over 16\pi G} (R-2\Lambda ) \ .    \eqno(2.1)
$$      
Motivated by the cosmological observation \cite{c1}-\cite{c2}, 
we take the case of a flat 
Friedmann-Lemaitre-Robertson-Walker metric, 
$$
ds^2=-dt^2+ a^2 (t) \sum_{k=1}^3 (dx^k)^2   \   .            \eqno(2.2)
$$
The action for this metric can be written as 
$$
S= \int \! {\d} t \  L \ , \qquad 
L={1 \over 16\pi G} (-6a \dot{a}^2  - 2 \Lambda a^3 ) \ ,    \eqno(2.3)
$$
where we neglected  an irrelevant constant. Here $a$ is the scalar factor of the universe. The constant we neglected does not effect the dynamics 
of the equations of motion from this Lagrangian, and hence can be neglected. 

Now  we  define unit such that   $ 8 \pi G = 1 $, and   
  obtain the Hamiltonian constraint as 
$$
{\cal H} = -{1 \over 12 a} p_a^2 +  \Lambda a^3 
\approx 0 \ ,                                                \eqno(2.4)
$$
where $p_a$ is the canonical momentum of $a$ . 
Now using the standard representation for the   canonical momentum, 
$$
p_a \rightarrow -i {{\d} \over {\d} a}    \ ,                \eqno(2.5)
$$
we obtain the Wheeler-DeWitt equation 
$$
\left[ {1 \over a^{p_o}} {{\d} \over {\d} a}
a^{p_o} {{\d} \over {\d} a} 
+ 12 \Lambda a^4 \right] \psi (a)  = 0 \ .             
$$
We can also write it as 
$$
\left[ {{\d}^2 \over {\d} a^2}
+ {p_o \over a} {{\d} \over {\d} a} 
+ 12 \Lambda a^4 \right] \psi (a)  = 0 \ .                   \eqno(2.6)
$$                                                
Here $p_o$ is the operator ordering parameter. The dependence of wave function of 
the universe on the cosmological constant has already been studied \cite{Hartle83}. 
In this paper, we will analyse the effect of operator ordering on the quantum fluctuations in quantum cosmology.

We consider $a$ as time. 
The Lagrangian for the third quantization which yields Eq. (2.6) is 
$$
{\cal L}_{3Q} = {1 \over 2}
\left[  a^{p_o} \left( {{\d}\psi(a) \over {\d} a}\right)^2
-12\Lambda a^{p_o +4} \psi (a)^2
\right] \ .                                                  \eqno(2.7)
$$ 
Note that if we define
$$
S_{3Q} = \int {\d}a \ {\cal L}_{3Q} \ ,                      \eqno(2.8)
$$
then we obtain Eq. (2.6) from 
$ \delta S_{3Q} = 0 $ . 

The canonical momentum for $\psi (a)$ is defined as 
$$
\pi (a) = {\del {\cal L}_{3Q} \over \del \left(
{{\d} \psi (a) \over {\d} a} \right)} 
= a^{p_o} {{\d} \psi (a) \over {\d} a} \ .                   \eqno(2.9)
$$
The Hamiltonian for the third quantization reads 
$$
\begin{array}{ll}
{\cal H}_{3Q} &= \pi (a) {{\d} \psi (a) \over {\d} a}
-{\cal L}_{3Q} \ , \\[5mm]
&={1 \over 2}\left[{1 \over a^{p_o}} \pi (a)^2 
+12\Lambda a^{p_o +4} \psi (a)^2
\right] \ . 
\end{array}                                                  \eqno(2.10)
$$

Now we   third quantize this theory  by imposing 
the equal time commutation relation as 
$$
[{\hat \psi} (a) , {\hat \pi} (a) ] = i \ ,                  \eqno(2.11)
$$
where hat represents an operator. 
Taking the $\rm Schr\ddot{o}dinger$ picture, we have 
the time-independent $c$-number $\psi$ for the operator 
${\hat \psi} (a)$ , 
so we can replace the operators as   
$$
{\hat \psi} (a) \rightarrow \psi \ , \qquad
{\hat \pi} (a) \rightarrow 
-i{\del \over \del \psi} \ .                                 \eqno(2.12)                             
$$ 
Then we obtain the $\rm Schr\ddot{o}dinger$ equation
$$
\begin{array}{ll}
&\dis{i{\del \Psi (a, \psi) \over \del a}} = {\hat {\cal H}}_{3Q} 
\Psi (a, \psi) \ , \\[5mm]
&\qquad\ \ {\hat {\cal H}}_{3Q}= 
\dis{{1 \over 2}\left[- {1 \over a^{p_o}} 
{\del^2 \over \del \psi^2}  
+12\Lambda a^{p_o +4} \psi^2
\right]} \ , 
\end{array}                                                  \eqno(2.13)
$$
where $\Psi (a, \psi )$ is the third quantized wave function of universes. 
This equation is the solution to a third quantized $\rm Schr\ddot{o}dinger$ equation, 
and hence it describes a muti-universe state, just as the solution to a second quantized 
$\rm Schr\ddot{o}dinger$ equation would describe a multi-particle state.

\section{Uncertainty Relation}\label{2}
It may be noted  that the universe is expected to be formed  from 
quantum fluctuations in the third quantized formalism \cite{ab}. 
Thus, we expect that, at the 
 beginning,  the universe will be dominated by quantum fluctuations.   
However, it is known that  the universe is described by a classical geometry 
at later stages. Thus, we expect that in the third quantized formalism, the early stages of the 
universe should be dominated by quantum fluctuations, and the later times  the universe should 
be described by a classical geometry. By classical geometry we mean that 
the quantum fluctuations of the geometry of 
the universe should get minimized. We can also make the definition of early and later times for the universe 
more precise by defining the early times by the limit $a \rightarrow 0$, 
and the later times by the limit $a \rightarrow \infty$.

Now we will analyse the uncertainty relation for the universe to analyse the behavior of quantum fluctuations 
at different stages of the cosmic expansion. In order to do that, we  
 assume that the solution to Eq. (2.13) has the Gaussian form 
$$
\Psi (a, \psi) = C {\rm exp} \left\{ -{1 \over 2}A(a)
[\psi-\eta (a)]^2 +i B(a)[\psi-\eta (a)]
\right\} \ ,                                                 \eqno(3.1)
$$
where $C$ is a constant, $A(a)=D(a)+iI(a)$, and $A(a), B(a), \eta (a)$ 
should be determined from Eq. (2.13). 
The inner product of two third quantized wave functions 
$\Psi_1$ and $\Psi_2$ can be defined as 
$$
\langle \Psi_1 , \Psi_2 \rangle 
=\int \! d \psi \, \Psi_1^*(a,\psi)
 \Psi_2(a,\psi)          .                                   \eqno(3.2)
$$
Let us calculate Heisenberg's uncertainty relation. 
The dispersion of $\psi$ is defined as
$$
(\Delta \psi)^2 \equiv \langle \psi^2 \rangle
-\langle \psi \rangle^2 \ , \qquad
\langle \psi^2  \rangle 
= {\langle \Psi ,  \psi^2 \Psi \rangle \over 
\langle \Psi , \Psi \rangle } \ .                            \eqno(3.3)
$$
Using Eqs. (3.1), (3.2) and (3.3), we have \cite{f}-\cite{f1}
$$
\langle \psi^2  \rangle = {1 \over 2D(a)} 
+ \eta^2 (a) \ , 
\quad \langle \psi \rangle = \eta (a) \ , \quad {\rm and} \quad
(\Delta \psi)^2 = {1 \over 2D(a)} \ .                        \eqno(3.4)
$$
The dispersion of $\pi$ is defined as
$$
(\Delta \pi)^2 \equiv \langle \pi^2 \rangle
-\langle \pi \rangle^2 \ , \qquad
\langle \pi^2  \rangle 
= {\langle \Psi ,  \pi^2 \Psi \rangle \over 
\langle \Psi , \Psi \rangle } \ .                            \eqno(3.5)
$$
Then we obtain 
$$
\begin{array}{ll}
&\dis{\langle \pi^2 \rangle 
= {D(a) \over 2}+{I^2 (a) \over 2D(a)}
+B^2 (a) , 
\quad \langle \pi \rangle = B(a)} \ ,  \\[5mm]     
{\rm and} \qquad \qquad \qquad \quad 
&\dis{(\Delta \pi)^2 
= {D(a) \over 2}+{I^2 (a) \over 2D(a)}} \ . 
\end{array}                                                  \eqno(3.6)
$$
Therefore the uncertainty  can be written as
$$
(\Delta \psi)^2 (\Delta \pi)^2
={1 \over 4} \Biggl( 1+ {I^2 (a) \over D^2 (a)} 
\Biggr) \  .                                                 \eqno(3.7)
$$

Substituting the assumption (3.1) to Eq. (2.13), 
we obtain the equation for $A(a)$ 
as 
$$
-{i \over 2}{{\d} A(a) \over {\d} a}
=-{1 \over 2 a^{p_o}} A(a)^2 
+ 6 \Lambda a^{p_o +4} \ .                                   \eqno(3.8)
$$
(We obtain three equations for $A(a), B(a), \eta (a)$ by 
comparing the order of $\psi$ in Eq. (2.13), 
but Eq. (3.8) is enough 
for the discussion of the Heisenberg uncertainty relation.)
Defining 
$$
\sigma \equiv a^{1-p_o} \ ,                                  \eqno(3.9)
$$
we obtain 
$$
-i{1-p_o  \over 2} {{\d} A(\sigma) \over {\d} \sigma}
+{A(\sigma)^2 \over 2}
-6 \Lambda \sigma^{2p_o +4 \over 1-p_o }
=0 \ .                                                      \eqno(3.10)
$$
Here we have assumed $p_o \neq 1$ . 
Let us define a function $u(\sigma)$ by the equation, 
$$
A(\sigma) = -i(1-p_o) 
{{\d} \ {\rm ln} \ u(\sigma) \over {\d} \sigma} \ .         \eqno(3.11)
$$
Then we have 
$$
{{\d}^2 u(\sigma) \over {\d} \sigma^2}
+{12 \Lambda \over (1-p_o)^2}\sigma^{2p_o +4 \over 1-p_o } 
 u(\sigma) 
= 0 \ .                                                     \eqno(3.12)
$$
This equation can be solved using a Bessel function as 
$$
u(\sigma) = \sigma^{1 \over 2} {\cal B}_{1-p_o \over 6}
\left(
2\sqrt{{\Lambda \over 3}}  \sigma^{3 \over 1-p_o}
\right) \ ,                                                 \eqno(3.13)
$$
where ${\cal B}$ is a Bessel function which satisfies \cite{Abramowitz-Stegun}
$$
{{\d}^2 {\cal B}_{1-p_o \over 6}(z) \over {\d} z^2}
+{1 \over z}{{\d} {\cal B}_{1-p_o \over 6}(z) \over {\d} z}
+\left( 1-{\bigl( {1-p_o \over 6} \bigr)^2 \over z^2}
\right) 
{\cal B}_{1-p_o \over 6}(z) =0 \ .                          \eqno(3.14)
$$
Therefore we obtain the general solution to eq. (3.12) as 
$$
u(\sigma) = c_J \sigma^{1 \over 2} J_{1-p_o \over 6}
\left( 2\sqrt{{\Lambda \over 3}}  
\sigma^{3 \over 1-p_o} \right)
+ c_Y \sigma^{1 \over 2} Y_{1-p_o \over 6}
\left( 2\sqrt{{\Lambda \over 3}}  
\sigma^{3 \over 1-p_o}  \right) 
\ ,                                                        \eqno(3.15)
$$
where $c_J$ and $c_Y$ are arbitrary constants and 
$J_{1-p_o \over 6}$ and $Y_{1-p_o \over 6}$ are Bessel functions.  

Now if we define 
$$
z \equiv {2 \sqrt{{ \Lambda} \over 3}} \sigma^{3 \over 1-p_o}
= {2 \sqrt{{ \Lambda} \over 3}} a^3 \ .                  \eqno(3.16)
$$
So, Eq. (3.15) can be written as  
$$
u(z) = \biggl( {z \over 2 \sqrt{{ \Lambda} \over 3}} 
\biggr)^{1-p_o \over 6} 
[ c_J J_{1-p_o \over 6} (z) + c_Y Y_{1-p_o \over 6} (z) ] 
\ .                                                        \eqno(3.17)
$$
Now from Eqs. (3.11), (3.16),  we obtain  
$$
\begin{array}{ll}
A(z) &= \dis{-i(1-p_o) {{\d} z \over {\d} \sigma} 
{{\d}\ {\rm ln} \ u(z) \over {\d} z} }\\[5mm]
&= \dis{-i \, 6\sqrt{\Lambda \over 3}
\left( {z \over 2\sqrt{\Lambda \over 3}} \right)^{p_o+2 \over 3}
{c_J J_{-5-p_o \over 6} (z) + c_Y Y_{-5-p_o \over 6} (z) \over 
c_J J_{1-p_o \over 6} (z) + c_Y Y_{1-p_o \over 6} (z)} }
\ , 
\end{array}                                                \eqno(3.18)
$$
where we have used \cite{Abramowitz-Stegun} 
$$
{{\d} {\cal B}_{1-p_o \over 6} (z) \over {\d} z} 
= {\cal B}_{-5-p_o \over 6} (z) 
-{1-p_o \over 6z}{\cal B}_{1-p_o \over 6} (z) \ .          \eqno(3.19) 
$$

As $A(z) = D(z) + i I(z)$ , we find 
(note that $c_J c^*_Y - c^*_J c_Y$ is a pure imaginary number)   
$$
D(z)=\dis{{i \, 6\sqrt{\Lambda \over 3}
\left( {z \over 2\sqrt{\Lambda \over 3}} \right)^{p_o+2 \over 3} 
\over 
\pi z \vert c_J J_{1-p_o \over 6} (z) 
+ c_Y Y_{1-p_o \over 6} (z) \vert^2}
(c_J c^*_Y - c^*_J c_Y)} \ .                                \eqno(3.20)
$$
Here we have used  \cite{Abramowitz-Stegun}
$$
J_{1-p_o \over 6} (z) Y_{-5-p_o \over 6} (z)
-J_{-5-p_o \over 6} (z) Y_{1-p_o \over 6} (z)
= {2 \over \pi z} \ ,                                      \eqno(3.21)
$$
and 
$$
\begin{array}{ll}
I(z)=&-\dis{3\sqrt{\Lambda \over 3} 
\left( {z \over 2\sqrt{\Lambda \over 3}} \right)^{p_o+2 \over 3}
\over   
\vert c_J J_{1-p_o \over 6} (z) 
+ c_Y Y_{1-p_o \over 6} (z) \vert^2} \\[6mm]
&\times 
\biggl[ 2\vert c_J \vert^2 J_{-5-p_o \over 6}(z) J_{1-p_o \over 6}(z)
+2\vert c_Y \vert^2 Y_{-5-p_o \over 6}(z)Y_{1-p_o \over 6}(z)  \\[3mm]
&\quad +(c_J c_Y^* + c_J^* c_Y)
\Bigl( J_{-5-p_o \over 6}(z) Y_{1-p_o \over 6}(z) \\[3mm]
&\qquad \qquad \qquad \quad \quad
+ J_{1-p_o \over 6}(z) Y_{-5-p_o \over 6}(z) \Bigr)
 \biggr] \ .                          
\end{array}                                                \eqno(3.22)
$$
Therefore if we assume $c_J c^*_Y - c^*_J c_Y \neq 0$  
(note that in this case both of $c_J ,  c_Y$ are nonzero), 
we obtain 
$$
\begin{array}{ll}
\dis{I(z)^2 \over D(z)^2}
=&-\dis{\pi^2 z^2 \over 4 (c_J c^*_Y - c^*_J c_Y)^2} \\[6mm]
&\times
\biggl[ 2\vert c_J \vert^2 J_{-5-p_o \over 6}(z) J_{1-p_o \over 6}(z)
+2\vert c_Y \vert^2 Y_{-5-p_o \over 6}(z)Y_{1-p_o \over 6}(z)  \\[3mm]
&\quad +(c_J c_Y^* + c_J^* c_Y)
\Bigl( J_{-5-p_o \over 6}(z) Y_{1-p_o \over 6}(z) \\[3mm]
&\qquad \qquad \qquad \quad \quad
+ J_{1-p_o \over 6}(z) Y_{-5-p_o \over 6}(z) \Bigr)
 \biggr]^2 \ . 
\end{array}                                                \eqno(3.23)
$$
Substituting Eq. (3.23) to Eq. (3.7), 
we obtain the Heisenberg uncertainty relation. 
This uncertainty relation can be used to analyse the   
behavior of the geometry during the cosmic expansion.

\section{Operator Ordering}\label{b}
In this section, we will discuss the operator ordering for the third quantization 
of the model studied in the previous sections. We will 
 estimate the order of the uncertainty relation both 
at the late and early times during the cosmic expansion.

The late times for the cosmic expansion can be defined as     $a \rightarrow \infty$ i. e.,  
$z \rightarrow \infty$ from Eq. (3.16). Now we can write 
\cite{Abramowitz-Stegun}
$$
J_{\nu} (z) \sim \sqrt{ 2 \over \pi z} 
\cos \left( z-{\nu \pi \over 2} - {\pi \over 4} \right) \ , 
\quad 
Y_{\nu} (z) \sim \sqrt{ 2 \over \pi z} 
\sin \left( z-{\nu \pi \over 2} - {\pi \over 4} \right) 
\ ,                                                        \eqno(4.1)
$$
where $\nu = {-5-p_o \over 6} \ {\rm and} \ {1-p_o \over 6}$ , 
so we obtain from Eq. (3.23) 
$$
\begin{array}{ll}
\dis{I(z)^2 \over D(z)^2}
&\sim -\dis{1 \over  (c_J c^*_Y - c^*_J c_Y)^2} \\[6mm]
&\qquad\times
\biggl[ 2\vert c_J \vert^2 
\cos \Bigl( z + {p_o + 2 \over 12}\pi \Bigr)
\cos \Bigl( z + {p_o -4 \over 12}\pi \Bigr) \\[5mm]
&\qquad \ \, 
+2\vert c_Y \vert^2 \sin \Bigl( z + {p_o + 2 \over 12}\pi \Bigr)
\sin \Bigl( z + {p_o -4 \over 12}\pi \Bigr)  \\[5mm]
&\qquad \ \, 
+(c_J c_Y^* + c_J^* c_Y)
\sin \Bigl( 2z + {p_o -1 \over 6} \pi \Bigr)
 \biggr]^2 \\[5mm]
&\sim O(1)    \ . 
\end{array}                                                \eqno(4.2)
$$ 
This along with Eq. (3.7) indicate  that at late times, i. e., in the limit 
$a \rightarrow \infty$,     the spacetime 
 can become classical in the sense 
that the quantum fluctuations become minimum.

On the other hand at early times namely when 
$a \rightarrow 0$ i.e.,  
$z \rightarrow 0$ from Eq. (3.16), we must 
divide the cases by the value of the operator ordering parameter 
$p_o$ . 

For example, when we choose as the usual case 
$$
p_o = -1  , \                                              \eqno(4.3)
$$  
we have  for early times  \cite{Abramowitz-Stegun} 
$$
\begin{array}{ll}
\dis{ J_{1 \over 3}(z)} 
&\sim \dis{ {1 \over \Gamma \Bigl( {4 \over 3} \Bigr)}
 \left( {z \over 2} \right)^{1 \over 3} ,
\qquad J_{-{2 \over 3}}(z) 
\sim {1 \over \Gamma \Bigl( {1 \over 3} \Bigr)}
\left({z \over 2} \right)^{-{2 \over 3}}
 },  \\[6mm]
\dis{Y_{1 \over 3}(z)} 
&\sim \dis{ -{1 \over \pi} \Gamma \biggl({1 \over 3} \biggr) 
\biggl( {z \over 2} \biggr)^{-{1 \over 3}} , 
\qquad Y_{-{2 \over 3}}(z) 
\sim {1 \over \sqrt{3} \Gamma \Bigl( {1 \over 3}\Bigr)} 
\left({z \over 2} \right)^{-{2 \over 3}}} .  
\end{array}                                                \eqno(4.4)                         
$$
So, we  can use Eq. (3.23) to obtain 
$$
\begin{array}{ll}
\dis{I(z)^2 \over D(z)^2} &\sim 
-\dis{{1 \over  (c_J c^*_Y - c^*_J c_Y)^2}
\biggl[ {2 \over \sqrt{3}} \vert c_Y \vert^2
+ (c_J c_Y^* + c_J^* c_Y)
\biggr]^2} \\[6mm]
&\sim O(1) \ . 
\end{array}                                                \eqno(4.5)
$$
This along with  Eq. (3.7) indicate  that at early times, i. e., in the limit 
 $a \rightarrow 0$,   the spacetime can again  become classical. Thus, this value of the operator ordering does not seem to be 
 physically plausible as the spacetime cannot become classical at early times, and we do expect that at early times the 
 quantum fluctuations should dominate. 

On other hand,  when we choose 
$$
p_o = -5 , \                                               \eqno(4.6)
$$
we have for   early times  \cite{Abramowitz-Stegun}
$$
\begin{array}{ll}
\dis J_0(z) &\sim \dis{1-{z^2 \over 4}} ,
\qquad J_1(z) \sim \dis{{z \over 2}} ,  \\[5mm]
Y_0(z) &\sim \dis{{2 \over \pi} \ln z}  , \ 
\qquad Y_1(z) \sim \dis{-{2 \over \pi z}}. 
\end{array}                                                \eqno(4.7)                   
$$
So, we can write  
$$
{I^2 (z) \over D^2 (z)} \sim 
-{16 \vert c_Y \vert^4 \over 
\pi^2 (c_J c_Y^* -c_J^* c_Y)^2}
( \ln z )^2 \sim \infty  .                                 \eqno(4.8)
$$
This along with  Eq.(3.7) indicate  that the fluctuation of the  universe 
field becomes very large at early times, i. e., in the limit  
$a \rightarrow 0$ in the third quantized formalism.  
Therefore the quantum  fluctuations dominate the structure of spacetime  for the small values of the 
scale factor of the universe. Thus, it seems to be the physically plausible result as we do expect the 
quantum fluctuations to dominate the universe at the very early stages. 
It may be noted that a similar result has been discussed in the context of 
  $f(R)$ gravity \cite{f}-\cite{f1}. Here we have seen that the physics of the system might depend 
  critically on operator ordering. This can be used to put a constraint on the operator ordering parameter, 
  allowing only those values which give physically expected results.

\section{Another Model for the Universe}\label{a}
It is important to check if the observations of the previous sections were a specific feature of the particular
minisuperspace model we analysed. Thus, it is important to perform a similar analysis for a different 
model for the universe. So, in 
  this section,  we analyse  another  model for the universe. We will study the third quantization of a 
   closed universe which is filled with     
constant vacuum energy density and  radiation. 
The Wheeler-DeWitt equation for this model can be written as \cite{d}
$$
\left[ {{\d}^2 \over {\d} a^2}
+ {p_o \over a} {{\d} \over {\d} a} 
- k_2 a^2 + k_4 \rho_v a^4 + k_0 \epsilon
 \right] \psi (a)  = 0 \ .                                 \eqno(5.1)
$$                                                
Here $a$ is again the scale factor of the universe, 
$p_o$ is the operator ordering parameter, and \cite{d}
$$ 
k_2={9 \pi^2 \over 4G^2 \hbar^2} \ , \ 
k_4={6 \pi^3 \over G \hbar^2} \ ,  \ 
k_0={6 \pi^3 \over G \hbar^2} 
\ .  \                                                      \eqno(5.2)
$$   
The behavior of the wave function of the universe, and its dependence on the vacuum energy and radiation 
has been studied \cite{fd}. 
We have denoted   the vacuum energy density by 
$\rho_v$, and the     radiation by 
$\epsilon$. Now repeating the analysis performed for the previous model, we 
  obtain the $\rm Schr\ddot{o}dinger$ equation for this model, 
$$
\begin{array}{ll}
&\dis{i{\del \Psi (a, \psi) \over \del a}} = {\hat {\cal H}}_{3Q} 
\Psi (a, \psi) \ , \\[5mm]
&\qquad\ \ {\hat {\cal H}}_{3Q}= 
\dis{{1 \over 2}\left[- {1 \over a^{p_o}} 
{\del^2 \over \del \psi^2}  
+ a^{p_o} (- k_2 a^2 + k_4 \rho_v a^4 + k_0 \epsilon )
\psi^2
\right]} \ , 
\end{array}                                               \eqno(5.3)
$$
where $\Psi (a, \psi )$ is the third quantized 
wave function.    

If we assume that the solution to Eq. (5.3) is the same 
Gaussian form of Eq. (3.1), 
we obtain the equation for $A(a)$ as 
$$
-{i \over 2}{{\d} A(a) \over {\d} a}
=-{1 \over 2 a^{p_o}} A(a)^2 
+ {a^{p_o} \over 2} (- k_2 a^2 + k_4 \rho_v a^4 + k_0 \epsilon ) 
 \ .                                                      \eqno(5.4)
$$
Using the same equations as Eqs. (3.9) and (3.11), we have 
$$
{{\d}^2 u(\sigma) \over {\d} \sigma^2}
+{1 \over (1-p_o)^2}\left(
-k_2 \sigma^{2p_o +2 \over 1-p_o }
+k_4 \rho_v \sigma^{2p_o +4 \over 1-p_o }
+k_0 \epsilon \sigma^{2p_o \over 1-p_o } 
\right) u(\sigma) 
= 0 \ .                                                   \eqno(5.5)
$$
Though this equation is too complicated to solve exactly, 
 we need only the limiting cases for the  late times 
$a \rightarrow \infty$ and the early times 
$a \rightarrow 0$, so we will solve it in these limiting cases.
 
At the late times we obtain from Eqs. (5.4) and (5.5) 
$$
{{\d}^2 u(\sigma) \over {\d} \sigma^2}
+{1 \over (1-p_o)^2}
k_4 \rho_v \sigma^{2p_o +4 \over 1-p_o }
 u(\sigma) 
= 0 \ .                                                   \eqno(5.6)
$$
Since this equation is essentially the same one as Eq. (3.12), 
we obtain the same result that is at late times, i .e.,  in the limit   
$a \rightarrow \infty$,     the spacetime 
becomes classical in the sense 
that the quantum fluctuations get minimized.

At early times, we obtain from Eqs. (5.4) and (5.5)  
$$
{{\d}^2 u(\sigma) \over {\d} \sigma^2}
+{1 \over (1-p_o)^2}
k_0 \epsilon \sigma^{2p_o \over 1-p_o } 
 u(\sigma) 
= 0 \ .                                                   \eqno(5.7)
$$
The solution for this is given by \cite{Abramowitz-Stegun}, 
$$
u(\sigma) = c_J \sigma^{1 \over 2} J_{1-p_o \over 2}
\left( \sqrt{k_0 \epsilon}  
\sigma^{1 \over 1-p_o} \right)
+ c_Y \sigma^{1 \over 2} Y_{1-p_o \over 2}
\left( \sqrt{k_0 \epsilon}  
\sigma^{1 \over 1-p_o}  \right) 
\ ,                                                      \eqno(5.8)
$$
where $c_J$ and $c_Y$ are arbitrary constants.  
If we define 
$$
z \equiv \sqrt{k_0 \epsilon} \, \sigma^{1 \over 1-p_o}
= \sqrt{k_0 \epsilon} \, a \ ,                             \eqno(5.9)
$$
then we have
$$
u(z) = \biggl( {z \over  \sqrt{k_0 \epsilon} }
\biggr)^{1-p_o \over 2} 
[ c_J J_{1-p_o \over 2} (z) + c_Y Y_{1-p_o \over 2} (z) ] 
\ .                                                      \eqno(5.10)
$$
So, from Eqs. (3.11), (5. 9), (5.10),  we obtain   \cite{Abramowitz-Stegun}
$$
A(z) 
= -i \sqrt{k_0 \epsilon}
\left( {z \over \sqrt{k_0 \epsilon}} \right)^{p_o}
{c_J J_{-1-p_o \over 2} (z) + c_Y Y_{-1-p_o \over 2} (z) \over 
c_J J_{1-p_o \over 2} (z) + c_Y Y_{1-p_o \over 2} (z)} 
\ .                                                     \eqno(5.11)
$$
Now repeating the analysis of the previous section, we obtain \cite{Abramowitz-Stegun}
$$
\begin{array}{ll}
\dis{I(z)^2 \over D(z)^2}
=&-\dis{\pi^2 z^2 \over 4 (c_J c^*_Y - c^*_J c_Y)^2} \\[6mm]
&\times
\biggl[ 2\vert c_J \vert^2 J_{-1-p_o \over 2}(z) J_{1-p_o \over 2}(z)
+2\vert c_Y \vert^2 Y_{-1-p_o \over 2}(z)Y_{1-p_o \over 2}(z)  \\[3mm]
&\quad +(c_J c_Y^* + c_J^* c_Y)
\Bigl( J_{-1-p_o \over 2}(z) Y_{1-p_o \over 2}(z) \\[3mm] 
&\qquad \qquad \qquad \quad \quad
+ J_{1-p_o \over 2}(z) Y_{-1-p_o \over 2}(z) \Bigr)
 \biggr]^2 \ .  
\end{array}                                             \eqno(5.12)
$$
 
If we choose as the usual case
$$
p_o=-1 , \                                              \eqno(5.13)
$$
we have the same relation as (4.8). 
So,  this along with  Eq.(3.7) indicate  that the fluctuation of 
the third quantized universe 
field becomes large at early times namely 
$a \rightarrow 0$. 
Therefore the quantum effects dominate for the small 
values of the scale factor of the universe. 

However if we choose other operator ordering parameter 
for example 
$$
p_o=0 , \                                               \eqno(5.14)
$$
at early times,  we have \cite{Abramowitz-Stegun} 
$$
\begin{array}{ll}
\dis{ J_{1 \over 2}(z)} 
&\sim \dis{ {2 \over \sqrt{\pi}}
 \left( {z \over 2} \right)^{1 \over 2} ,
\qquad J_{-{1 \over 2}}(z) 
\sim {1 \over \sqrt{\pi}}
\left({z \over 2} \right)^{-{1 \over 2}}
 },  \\[6mm]
\dis{Y_{1 \over 2}(z)} 
&\sim \dis{ -{1 \over \sqrt{\pi}}  
\biggl( {z \over 2} \biggr)^{-{1 \over 2}} , 
\qquad Y_{-{1 \over 2}}(z) 
\sim {2 \over \sqrt{\pi}} 
\left({z \over 2} \right)^{1 \over 2}}. 
\end{array}                                             \eqno(5.15)                         
$$
So, we can obtain from Eq. (5.12) 
$$
{I(z)^2 \over D(z)^2} \sim 
-{(c_J c_Y^* + c_J^* c_Y )^2 \over (c_J c_Y^* - c_J^* c_Y )^2}
\sim O(1) 
. \                                                     \eqno(5.16)
$$ 
This along with  Eq. (3.7) indicate  that at early times,  i. e.,    in the limit 
 $a \rightarrow 0$,  the spacetime becomes classical. This does not seem like a physical result. 
 Thus, again we have observed that the physics of this system depends on operator ordering, 
 and this can be used to constrain the form of the operator ordering. So, we have demonstrated 
 for two different models that the physical requirement that quantum fluctuations dominate the 
 early stages of the universe, and   spacetime becomes classical at later stages in the cosmic expansion, 
 holds for certain values of operator ordering. 
 
For the first model, we observe that $p_0 = -1$ is not physically plausible, and $p_0 = -5$ is physically acceptable value for the operator ordering. 
On the other hand, we observe that for the second model,  $p_0 = -1$ is physically acceptable, and $p_0 = 0$ is not physically acceptable value of the operator ordering. 
Thus, it seems that the exact value of the operator ordering chosen depends on the physical content of the universe. Hence, there seems to be no 
universal value for the operator ordering that can be chosen for all the physical problems, and rather a different value of the operator 
ordering has to be chosen for different models of quantum cosmology. However, the fact that some values of the operator ordering will give 
physically plausible results, and other values will give physically acceptable results, seems to be a general feature of all different models 
in quantum cosmology.  This is rather a surprising results, as one would have expected that any physical dependence on operator ordering 
should be a universal feature of all the models of quantum cosmology. However, in this paper, we have demonstrated that the physically acceptable 
value for the operator ordering to be model dependent. 

It may be noted that for both the   cosmological models,   we found that the quantum fluctuations 
depended on the exact value of the operator ordering chose. In the first model the universe was filled with a cosmological constant, 
and in the second model there was an important contribution coming from the radiation. So, the exact form of the Wheeler-DeWitt equation 
was different for both these models, and hence the wave function of the universe had a different form for both these models. 
However, it was interesting to note that in both these models the quantum fluctations in the early universe depended critically on the operator
ordering chosen. 
There is no mathematical way to prefer a specific form of operator ordering in the Wheeler-DeWitt equation. However, in this paper, we have 
demonstrated that operator ordering can have non-trivial physical consiquences. So, it is possible to use the physical requirement on the form 
of the wave function for the universe to prefer a specific form of the operator ordering. We do require the geometry of the universe to be dominated 
by quantum fluctuations at the very early stage, and this quantum state of universe is expected to give rise to a classical geometry 
at later stages. It has been demonstrated in this paper, that this requirement can be used to prefer a choice of operator ordering 
in the Wheeler-deWitt equation. However, the exact value of the operator ordering will also depend on the details of the cosmological  model being studied.

\section{Conclusion}\label{d}
In this paper, we studied the  operator ordering in the 
third quantized formalism. This was done by   analysing the  
the Wheeler-DeWitt equation in the third quantized formalism. Thus, an action was constructed such that 
 the Wheeler-DeWitt equation was obtained as its field equation. This action was then third quantized.
 We studied the 
 uncertainty relation for this third quantized    theory of gravity. We also discussed the operator order 
 for this theory. We   observed that the physical requirements are satisfied for 
 certain values of operator ordering. It is possible for  the universe to be formed from quantum 
 fluctuations in this third quantized formalism \cite{ab}.  So, it is expected that the 
 early stages of the universe will be dominated by quantum fluctuations. Furthermore, 
 it is known that the geometry  of the universe has to become classical at later stages of the cosmic expansion. 
 In this paper, we demonstrate that for certain 
 values of operator ordering parameter,  
 the quantum fluctuations do dominate the early state of the universe, and the spacetime 
 geometry does becomes classical,   
 at the later stages of the cosmic evolution. Hence, we can use this to put constraints on the form of operator 
 ordering chosen. It may be noted that we have demonstrate this to be the case for two different types of model.
However, the exact value of the operator ordering chosen depends on the details of the model being studied. 
We would like to point out that   
  factor ordering occur as there is an ambiguity in defining  two quantum operators   at the same point, and this is similar to 
the occurrence of an anomaly in quantum field theory. In fact, it has been demonstrated that the factor ordering in the Wheeler-DeWitt equation
can effect the classical Friedman equation \cite{d}. 
 
  It may be noted that the third quantization has also been studied in the context of  group field cosmology \cite{sg01}-\cite{sg02}. 
The FFBRST for the  group 
field cosmology has also been studied \cite{sg}-\cite{gs}. As recently lot of progress has been made on FFBRST \cite{01sg}-\cite{02sg}, it would be interesting 
to analyse the implications of these results on group field cosmology. 
 It will be also interesting to analyse  the uncertainty  in the context of   third 
quantized  Horava-Lifshitz     gravity  \cite{3}-\cite{4}.   This theory is based on the idea 
of modifying the scaling of space and time in such a way that the infrared limit of this theory 
coincides with general relativity. Thus,  Horava-Lifshitz     gravity is viewed as the 
 ultraviolet completion 
of   general relativity.  This theory uses the concept of  Lifshitz scaling from solid state physics, 
it is generally called 
 Horava-Lifshitz   theory $
t\rightarrow b^z t,\, {\bf x}\rightarrow b{\bf x}, 
$
where  $z$ is called  the dynamical critical exponent $z$, and we can assume that $z = 3$ \cite{3}-\cite{4}.
The Wheeler-DeWitt equation for the Horava-Lifshitz    gravity has been studied 
\cite{q1}-\cite{q}.  
The Wheeler-DeWitt equation for the Horava-Lifshitz  
gravity has also been used to study the cosmological constant problem \cite{q2}. 
The third quantization of the Horava-Lifshitz  gravity has also been studied \cite{mult}. 
This was done by associating the  eigenvalues   of the Wheeler-DeWitt equation
 with the cosmological constant. It will be interesting to study the uncertainty   for the 
 third quantized   Horava-Lifshitz  theory of gravity.

 The  Horava-Lifshitz  theory of gravity is closely related to the   
  gravity's rainbow \cite{rain}. 
  The deformation of   black holes   has been studied using 
  gravity's rainbow \cite{bh12}-\cite{hb12}. The Wheeler-DeWitt equation for 
 the  gravity's rainbow has already been studied using the
 second quantized formalism \cite{rain1}-\cite{rain2}. 
It will be interesting to perform the third quantization of the Wheeler-DeWitt equation for  gravity's rainbow.
This is because this third quantized Wheeler-DeWitt equation for  gravity's rainbow can be used for  
analysing different aspects of  virtual black holes. 
 It may be noted that a deformation of the kinetic part of the 
 Wheeler-DeWitt equation has been recently studied \cite{p1}-\cite{1p1q}. This 
 deformation of the Wheeler-DeWitt equation is based on generalized uncertainty principle \cite{p}-\cite{1p}.
 The third quantization 
 of this deformed Wheeler-DeWitt equation has also been discussed \cite{p2}. It will be interesting to study
 the uncertainty 
 in the context of the third quantization of a deformed Wheeler-DeWitt equation. 
 We can also discuss the operator 
 order for deformed   Wheeler-DeWitt equation. 
 It may be noted that it is possible 
to study this deformation of the Wheeler-DeWitt equation for the various modified theories of gravity like 
the $f(R)$ gravity and the  Horava-Lifshitz  gravity.
It will be interesting to study the effect of operator ordering for these  modifications to the Wheeler-DeWitt 
equation.

 \end{document}